# W-potentials in nonlinear biophysics of microtubules

**Slobodan Zdravković**[*]
Institut za Nuklearne Nauke Vinča, Laboratorija za Atomsku Fiziku (040),
11001 Beograd, Serbia

**Dragana Ranković**[†]
Farmaceutski fakultet, Univerzitet u Beogradu, 11221 Beograd, Serbia

ABSTRACT

In the present article we investigate the nonlinear dynamics of microtubules, the basic components of the eukaryotic cytoskeleton, and rely on the known general model. A crucial interaction among constitutive particles is modelled using W-potential. Three kinds of this potential are studied, symmetrical and two non-symmetrical. We demonstrate an advantage of the latter ones.

1. Introduction

There are two kinds of cells. These are eukaryotic, having a membrane-bound nucleus, and much simpler prokaryotic cells, without the nucleus. In the eukaryotes, an intracellular protein filament network exists. Microtubules (MTs), studied in this article, are the basic components of this cytoskeleton. They play essential roles in the shaping and maintenance of cells and in cell division. Also, MTs represent a traffic network for motor proteins moving along them.

Information about their structure and function can be found in many references [1-3]. Here, we mention some basic pieces of information only. MT is a long hollow cylinder spreading between the nucleus and cell membrane. Its surface is usually formed of 13 long structures called protofilaments (PFs), representing a series of heterodimers, as shown in Fig. 1. A key point is that the heterodimer, or dimer for short, is an electric dipole. This means that MT behaves as ferroelectric [4], which is crucial for many models of MTs. In this paper, we rely on the so-called general model (GM) [5].

The lengths of MTs vary from a few hundred nanometers up to meters in long nerve axons [6]. For most of the models, a dimer is a constitutive unit, which means that its internal structure is not taken into consideration. Its mass and length are $m = 1.8 \times 10^{-22}$ kg [4] and $l = 8$ nm [4,7,8], respectively. The longitudinal, tangential, and radial components of the electric dipole moment are $p_z = 1.13 \times 10^{-27}$ Cm, $p_\theta = 0.66 \times 10^{-27}$ Cm, and $p_r = -5.57 \times 10^{-27}$ Cm, respectively [9]. Hence, $p_z$ is in the direction of MT.

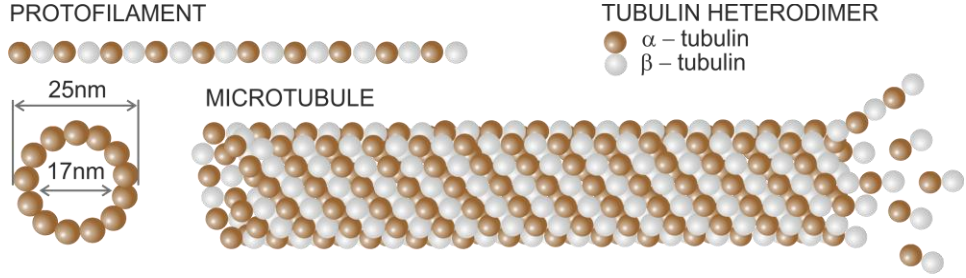

**Fig. 1.** Microtubule

In Section 2, we very briefly explain the used model. For one of the interactions between the dimers, we use W-potential energy, or potential for short. We study three types of them and obtain three dynamic equations of motion. These are crucial equations, solved in Section 3, while Section 4 is devoted to some concluding remarks.

## 2. W-potential within the general model of MTs

It was mentioned above that MT behaves like a ferroelectrics. Based on this fact is the first model [4] in a series of models describing MT nonlinear dynamics. As the used coordinate is a longitudinal one, the model belongs to the group of longitudinal models, as well as its improved ancestor [10]. There are a few degrees of freedom for each dimer, but the essential is the angular one. Hence, angular models represent a crucial group of models describing MT dynamics. The first in this series is the so-called u-model [11], a perhaps somewhat naive but rather important step in the evolution of the models. The next one is the general model (GM), mentioned above [5], which we rely on in this paper. We conclude the series with the model introduced recently [12]. This two-component angular model has been under active investigation and is not relevant for this work.

A starting point for all these models is Hamiltonian. For GM, it is [5]

$$H = \sum_n \left[ \frac{I}{2} \dot{\varphi}_n^2 + \frac{k}{2} (\varphi_{n+1} - \varphi_n)^2 + W(\varphi_n) - pE \cos \varphi_n \right], \tag{1}$$

where $I$ is a moment of inertia of the single dimer, $k$ is an inter-dimer stiffness parameter, $E > 0$ is the intrinsic electric field, and $p > 0$ stands for electric dipole moment. Notice that $E$ is the internal electric field, which means that a particular dimer exists in the field of all other dimers. The angle $\varphi_n$ describes the dimer`s oscillation and $n$ is its position. We recognize the kinetic and potential energies of the interaction of the two neighbouring dimers belonging to the same PF. A term $W(\varphi)$ represents the interaction of a single dimer with all other ones that do not belong to the same PF. It is called W-potential as it looks like a letter W, which will be clear later on. The very last term is coming from the fact that the electric dipole is in the field of all other ones. For this paper, the most important is the W-potential. We study the following three cases:

**Case 1.** W-potential is a symmetric function:

$$W_1 = -\frac{A}{2}\varphi_n^2 + \frac{B}{4}\varphi_n^4, \quad A > 0, \quad B > 0. \tag{2}$$

**Case 2.** W-potential is a non-symmetric function:

$$W_2 = -\frac{A}{2}\varphi_n^2 + \frac{B}{4}\varphi_n^4 - C\varphi_n. \tag{3}$$

**Case 3.** W-potential is a non-symmetric function:

$$W_3 = -\frac{A}{2}\varphi_n^2 + \frac{B}{4}\varphi_n^4 - \frac{D}{3}\varphi_n^3. \tag{4}$$

We use Hamilton's dynamical equation $I\ddot{\varphi}_n = -\partial H/\partial \varphi_n$, a continuum approximation $\varphi_n(t) \Rightarrow \varphi(x,t)$, series expansion of the cosine term, as well as $\varphi_{n\pm 1} \Rightarrow \varphi \pm \frac{\partial \varphi}{\partial x}l + \frac{1}{2}\frac{\partial^2 \varphi}{\partial x^2}l^2$ [10], and obtain the following dynamical equations of motion:

**Case 1.** $I\dfrac{\partial^2 \varphi}{\partial t^2} - kl^2 \dfrac{\partial^2 \varphi}{\partial x^2} - (A - pE)\varphi + \left(B - \dfrac{pE}{6}\right)\varphi^3 + \Gamma\dfrac{\partial \varphi}{\partial t} = 0,$ (5)

**Case 2.** $I\dfrac{\partial^2 \varphi}{\partial t^2} - kl^2 \dfrac{\partial^2 \varphi}{\partial x^2} - (A - pE)\varphi + \left(B - \dfrac{pE}{6}\right)\varphi^3 - C + \Gamma\dfrac{\partial \varphi}{\partial t} = 0,$ (6)

**Case 3.** $$I\frac{\partial^2\varphi}{\partial t^2} - kl^2\frac{\partial^2\varphi}{\partial x^2} - (A - pE)\varphi + \left(B - \frac{pE}{6}\right)\varphi^3 - D\varphi^2 + \Gamma\frac{\partial\varphi}{\partial t} = 0, \quad (7)$$

where $\Gamma$ is a viscosity parameter. Hence, we obtained partial differential equations. It is well known that, for a given wave equation, a travelling wave $\varphi(\xi)$ is a solution that depends upon $x$ and $t$ only through a unified variable

$$\xi = \kappa x - \omega t, \quad (8)$$

where $\kappa$ and $\omega$ are constants. According to Eqs. (5)-(8), we straightforwardly obtain ordinary differential equations (ODE):

**Case 1.** $$\alpha\psi'' - \rho\psi' - \psi + \psi^3 = 0, \quad (9)$$

**Case 2.** $$\alpha\psi'' - \rho\psi' - \psi + \psi^3 - \sigma = 0, \quad (10)$$

**Case 3.** $$\alpha\psi'' - \rho\psi' - \psi + \psi^3 - \delta\psi^2 = 0, \quad (11)$$

where,

$$\varphi = \sqrt{\frac{A - pE}{B - pE/6}}\psi \equiv K\psi, \quad \alpha = \frac{I\omega^2 - kl^2\kappa^2}{A - pE}, \quad \rho = \frac{\Gamma\omega}{A - pE}, \quad (12)$$

$$\sigma = \frac{C}{K(A - pE)}, \quad \delta = \frac{D}{\sqrt{(A - pE)\left(B - \frac{pE}{6}\right)}}, \quad (13)$$

and $\varphi' \equiv d\varphi/d\xi$. Experimental values of the involved parameters do not exist but our estimations strongly suggest $A > pE$ and $B > pE/6$ [5], which was used above.

Let us point out the great importance of the parameter $\alpha$. From Eqs. (1), (5)-(7), and (12), we conclude that its negative sign means that the elastic term is larger than the inertial one, and vice versa. Also, Eq. (12) can be written as

$$\alpha = \frac{I\omega^2 - kl^2\kappa^2}{A - pE} = \frac{I\kappa^2(\omega^2/\kappa^2 - kl^2/I)}{A - pE} \equiv \frac{I\kappa^2(v^2 - c^2)}{A - pE}, \quad (14)$$

where $v$ is the velocity of the solitary wave, while $c$ is the speed of sound. This means that the sign of $\alpha$ shows if the wave is subsonic or supersonic.

**3. Solutions of equations (9) – (11)**

There are many mathematical procedures for solving Eqs. (9)-(11). One of the simplest is a procedure that we call the tangent hyperbolic function method (THFM). According to THFM, we expect the solution $\psi$ as [13]

$$\psi = a_0 + \sum_{i=1}^{M}\left(a_i \Phi^i + b_i \Phi^{-i}\right), \qquad (15)$$

where $\Phi$ is the solution of the well-known Riccati equation

$$\Phi' = b + \Phi^2, \qquad (16)$$

and the parameters $a_0$, $a_i$, $b_i$, and $b$ should be determined. A solution of Eq. (16) depends on $b$ [10,13]. The one having physical sense is

$$\Phi = -\sqrt{-b}\,\tanh\!\left(\sqrt{-b}\,\xi\right), \qquad (17)$$

which holds for $b < 0$. The highest exponents are $\Phi^{M+2}$ and $\Phi^{3M}$, coming from $\psi''$ and $\psi^3$, respectively. This means that $M = 1$ in Eq. (15). Also, we set $b_i = 0$ because we are not interested in diverging solutions in this work, and Eq. (15) becomes.

$$\psi = a_0 + a\Phi. \qquad (18)$$

According to Eqs. (9)-(11), (16), and (18), we obtain the expression

$$K_3\Phi^3 + K_3'\Phi^{-3} + K_2\Phi^2 + K_2'\Phi^{-2} + K_1\Phi + K_1'\Phi^{-1} + K_0 = 0, \qquad (19)$$

which is satisfied if all the coefficients $K_i$ are simultaneously equal to zero. Of course, for $b_1 = 0$, the system is simplified. In what follows, we solve ODEs for all three cases, i.e., Eqs. (9)-(11).

**Case 1.**

This case was solved in Ref. [5], where the model we rely on was introduced. Using Mathematica, we easily obtain the following two solutions:

$$a_0^{(\pm)} = \pm\frac{1}{2}, \quad a^{(\pm)} = \pm\frac{2\rho}{3}, \quad b = -\frac{9}{16\rho^2}, \quad \alpha = -\frac{2\rho^2}{9}, \qquad (20)$$

which yields

$$\psi_1^{(\pm)} = \pm\frac{1}{2}\left[1 - \tanh\left(\frac{3}{4\rho}\xi\right)\right]. \qquad (21)$$

The functions $\psi_1^{(+)}$ and $\psi_1^{(-)}$ are kink and antikink solitons, respectively, or kinks for short. They are shown in Fig. 2 for $\rho = 1$.

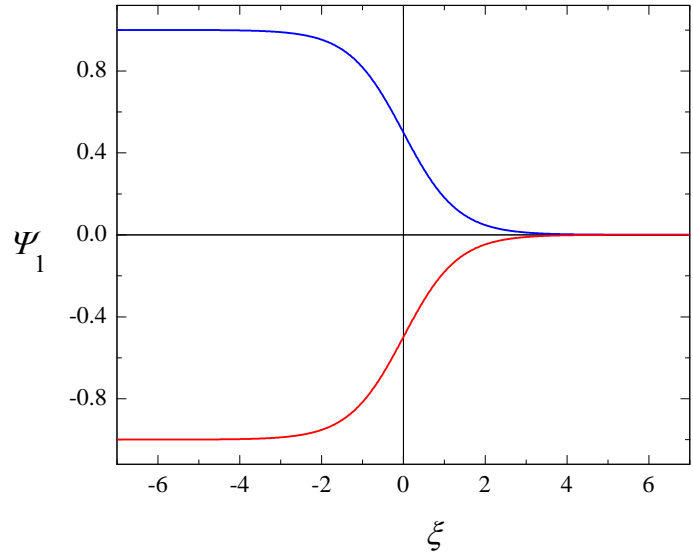

**Fig. 2.** Kink solitons $\psi_1^{(+)}$ (blue) and $\psi_1^{(-)}$ (red) for $\rho = 1$

If viscosity is neglected, that is for $\rho = 0$, the solution of Eq. (9) is

$$\psi_{10} = \tanh(\xi/a), \qquad \alpha = -a^2/2 < 0, \qquad (22)$$

where $a$ is an arbitrary constant introduced in Eq. (18). This function is similar to the one shown in Fig. 2 but it goes from -1 to 1, or from 1 to -1, depending on a sign of $a$.

AUTHORS

To obtain Eqs. (20)-(22), $A > pE$ and $B > pE/6$ were assumed, as mentioned above. The cases $(A - pE)\left(B - \dfrac{pE}{6}\right) < 0$, $A = pE$ and $B \neq \dfrac{pE}{6}$, and $B = pE/6$ were studied in Ref. [5]. No solution having physical sense, relevant for this article, was obtained.

**Case 2.**

Eq. (10) exists in Ref. [10], even though different models were established. Solutions, corresponding to Eq. (20), are [10]

$$b = \frac{3a_0^2 - 1}{a_1^2}, \quad \alpha = -\frac{a^2}{2}, \quad a = \frac{\rho}{3a_0}, \quad 8a_0^3 - 2a_0 + \sigma = 0. \tag{23}$$

Three real solutions of Eq. (23) are [10]

$$\left.\begin{aligned} a_{01} &= -\frac{1}{2\sqrt{3}}\left(\cos F + \sqrt{3}\sin F\right) \\ a_{02} &= \frac{1}{2\sqrt{3}}\left(-\cos F + \sqrt{3}\sin F\right) \\ a_{03} &= \frac{1}{\sqrt{3}}\cos F \end{aligned}\right\}, \tag{24}$$

where

$$F = \frac{1}{3}\arccos\left(\frac{\sigma}{\sigma_0}\right), \quad \sigma_0 = \frac{2}{3\sqrt{3}}. \tag{25}$$

Figure 3 shows how the functions $a_{0i}$ depend on $\sigma$. We notice that $a_{01}^2 < 1/3$. Also, according to Eqs. (24) and (25), we see that $a_{03}(\sigma_0) = 1/\sqrt{3}$. This means that three real roots of Eq. (23) exist for $\sigma < \sigma_0$, as $b < 0$. There is one real solution for $\sigma > \sigma_0$ [10]. However, the requirement $\sigma > \sigma_0$ is nothing but $3a_0^2 > 1$. This case corresponds to the positive $b$, which brings about the diverging solution [10].

The finite solutions, i.e., the functions $\psi_{2i}$, are determined by Eqs. (17), (18), and (23)-(25). They are

$$\psi_{2i}(\xi) = a_{0i} - \sqrt{1 - 3a_{0i}^2}\, \tanh\left(\frac{3a_{0i}}{\rho}\sqrt{1 - 3a_{0i}^2}\, \xi\right), \quad i = 1, 2, 3, \qquad (26)$$

shown in Fig. 4.

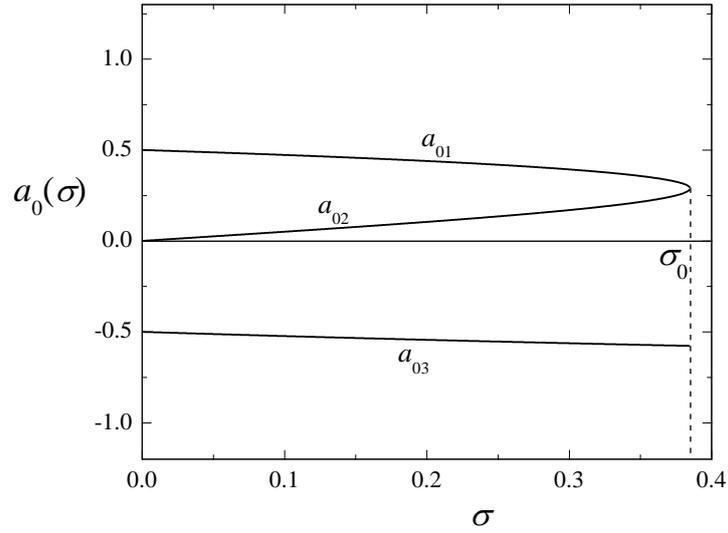

**Fig. 3.** The parameters $a_{0i}$ as functions of $\sigma$

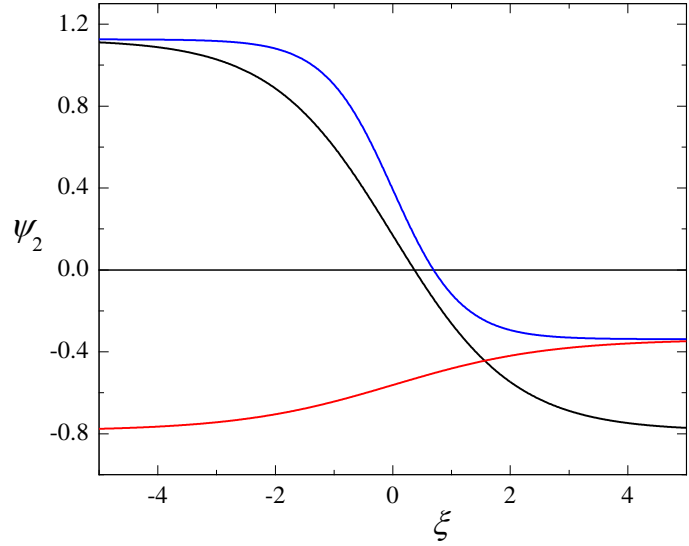

**Fig. 4.** Kink solitons $\psi_{21}$ (blue), $\psi_{22}$ (red), and $\psi_{23}$ (black) for $\rho = 1$ and $\sigma = 0.3$

According to Eqs. (13) and (23), we see that the approximation $\rho = 0$ yields to $a_0 = 0$ and $\sigma = 0$ and, consequently, to the symmetric potential.

**Case 3.**

It was explained above that Eq. (19) gives a system of equations from which we obtain values of the parameters determining the function $\psi$. As we study the case $a_1 \equiv a \neq 0$ and $b_1 = 0$, the coefficients $K_1'$, $K_2'$, and $K_3'$ disappear. Hence, we use Eqs. (11), (16), and (18) and obtain Eq. (19) for this particular case. The coefficients for $\phi^3$, $\phi^2$, $\phi^1$, and $\phi^0 = 1$ respectively bring about the following system

$$\left.\begin{array}{l} 2\alpha + a^2 = 0 \\ 3a_0 a - \rho - \delta a = 0 \\ 2\alpha b - 1 + 3a_0^2 - 2\delta a_0 = 0 \\ \rho ab + a_0 - a_0^3 + \delta a_0^2 = 0 \end{array}\right\}. \qquad (27)$$

According to Eq. (27), we obtain $\alpha = -a^2/2$, like in the previous two cases, and the three solutions for $a_0$, $a$, and $b$. These solutions are

$$a_0^{(1)} = \frac{\delta}{2}, \quad a^{(1)} = \frac{2\rho}{\delta}, \quad b^{(1)} = -\frac{\delta^2 K_0^2}{16\rho^2}, \qquad (28)$$

$$a_0^{(2)} = \frac{\delta - K_0}{4}, \quad a^{(2)} = -\frac{4\rho}{\delta + 3K_0}, \quad b^{(2)} = -\frac{(\delta + 3K_0)^2}{32\rho^2(\delta^2 + \delta K_0 + 2)} \qquad (29)$$

$$a_0^{(3)} = \frac{\delta + K_0}{4}, \quad a^{(3)} = -\frac{4\rho}{\delta - 3K_0}, \quad b^{(3)} = -\frac{(\delta - 3K_0)^2(\delta^2 + \delta K_0 + 2)}{128\rho^2}, \qquad (30)$$

where

$$K_0 = \sqrt{\delta^2 + 4}. \qquad (31)$$

It may be useful to keep in mind that the different analytical procedures, or different kinds of software, can bring about the expressions for the parameter that looks different from those given by Eqs. (29) and (30). For example, the alternative expressions for $b^{(3)}$ could be

$$b^{(3)} = -\frac{18 + 8\delta^2 + \delta^4 + 6\delta K_0 + \delta^3 K_0}{32\rho^2} \qquad (32)$$

and

$$b^{(3)} = -\frac{(2\delta^2 + 9)^2 [2\delta^4 + 10\delta^2 + 9 + 2\delta(\delta^2 + 3)K_0]}{\rho^2 (\delta + 3K_0)^4}. \qquad (33)$$

Therefore, all the three expressions are identical, which might not be obvious at first glance.

The final solutions, i.e., the functions $\psi_3^{(1)}$, $\psi_3^{(2)}$, and $\psi_3^{(3)}$, can be obtained according to Eqs. (17), (18), and (28)-(31). They are

$$\psi_{31}(\xi) = \frac{\delta}{2} - \frac{K_0}{2} \tanh\left(\frac{\delta K_0}{4\rho}\xi\right), \qquad (34)$$

$$\psi_{32}(\xi) = \frac{\delta - K_0}{4} + \frac{1}{\Lambda} \tanh\left(\frac{\delta + 3K_0}{4\rho\Lambda}\xi\right), \qquad (35)$$

$$\psi_{33}(\xi) = \frac{\delta + K_0}{4} + \frac{\Lambda}{4} \tanh\left(\frac{(\delta - 3K_0)\Lambda}{16\rho}\xi\right), \qquad (36)$$

where

$$\Lambda = \sqrt{2\delta^2 + 2\delta K_0 + 4}. \qquad (37)$$

They are shown in Fig. 5 for $\delta = 0.3$.

## 4. Discussions and future research

Each of the three potentials given by Eqs. (2)-(4) has two minima and one maximum. In terms of the functions $\psi$, we can talk of the right and left minima, and of the maximum, denoted as $\psi_R$, $\psi_L$, and $\psi_M$, respectively. Let us compare Figs. 2 and 4. The solutions in Fig. 2 represent the transitions $\psi_R \to \psi_M$ (blue) and $\psi_L \to \psi_M$ (red). Both transitions can be seen in Fig. 4, but there is one more. This is $\psi_R \to \psi_L$ (black), representing a transition from a deeper minimum to a shallower one. An obvious conclusion is that the non-symmetric potential $W_2$ is better than the symmetric one. Also, the same conclusion is suggested by the geometry of MTs.

The potentials $W_2$ and $W_3$ bring about Figs. 4 and 5. These figures are basically of the same value, i.e., describe all three transitions. This means that it is difficult to state which potential is better for MT modelling. In the case of $W_3$, the transition $\psi_R \to \psi_L$ goes slowly in comparison to $W_2$ (black lines), but this depends on the values of the parameters $\sigma$ and $\delta$. This might mean that $W_3$ is better but we are not ready for any suggestions without further research. We should notice that $\psi_M = 0$ in cases $W_1$ and $W_3$. This indicates that the unstable orientations of the dimers are exactly in the direction of PF. Unfortunately, the orientation of the dimers has not been experimentally determined yet. We can state that the potentials $W_2$ and $W_3$ are more convenient for MT modelling than the symmetric one. This issue should be studied within a new model [12].

The equations (5)-(7) were solved using the continuum approximation. The question of whether MT is a discrete or continuum system was studied in Ref. [14].

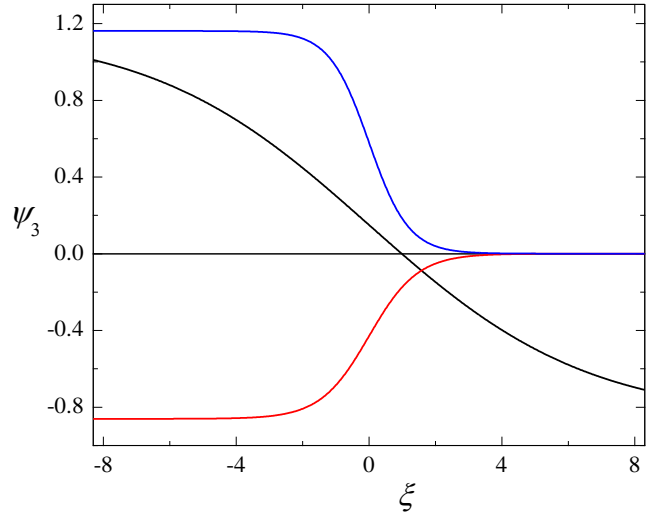

**Fig. 5.** Kink solitons $\psi_{31}$ (blue), $\psi_{32}$ (red), and $\psi_{33}$ (black) for $\rho = 1$ and $\delta = 0.3$

---

* e-mail address: szdjidji@vin.bg.ac.rs
† e-mail address: dragana.rankovic@pharmacy.bg.ac.rs